\documentclass[english]{article}
\usepackage{fontenc}
\usepackage[latin1]{inputenc}
\usepackage{amsmath}
\usepackage{fontenc}
\usepackage{babel}

\input{tcilatex}

\begin{document}

\title{\textbf{Effective energy-momentum tensor of strong-field QED with
unstable vacuum}}
\author{S.P. Gavrilov\thanks{%
E-mail: gavrilovsergeyp@yahoo.com} \\
\\
Department of general and experimental physics,\\
Herzen State Pedagogical University of Russia, \\
Moyka emb. 48, 191186 St. Petersburg, Russia}
\maketitle

\begin{abstract}
We study the influence of a vacuum instability on the effective
energy-momentum tensor (EMT) of QED, in the presence of a quasiconstant
external electric field, by means of the relevant Green functions. In the
case when the initial vacuum, $|0,in>$, differs essentially from the final
vacuum, $|0,out>$, we find explicitly and compared both the vacuum average
value of EMT, $<0,in|T_{\mu \nu }|0,in>$, and the matrix element, $%
<0,out|T_{\mu \nu }|0,in>$. In the course of the calculation we solve the
problem of the special divergences connected with infinite time T of acting
of the constant electric field. The EMT of pair created by an electric field
from the initial vacuum is presented. The relations of the obtained
expressions to the Euler-Heisenberg's effective action are established.

PACS numbers: 12.20.Ds, 11.10.Wx
\end{abstract}

\section{Introduction}

Currently, an effective action method, originating with Euler and
Heisenberg's one-loop effective action, is one of the commonly used
approaches of QFT. Nevertheless, we can see that if an external electric
field is involved then naive calculations by analogy with a magnetic field
case can be erroneous. For example, thermally influenced pair production in
a constant electric field has been searched via several attempts of
generalization for one-loop effective action at finite temperature with
extremely contrary results. We would like to express that the vacuum
instability in an electric background opens additional channels of
interaction due to particles creation from the vacuum. That is the reason
that the above mentioned simple analogy does not work, regardless of thermal
influence, and we have to refine setting up a problem. For simplicity of
explanation in this talk we suppose that the temperature is equal to zero%
\footnote{%
We will present the relevant generalization for one-loop effects at finite
temperature anywhere.}.

The relevant intense field method, applicable to the theory with unstable
vacuum in the case of a time-varying external field (called generalized
Furry representation), can be found in a book \cite{FGS91}\footnote{%
The extension of such an approach for finite temperature QED was presented
in \cite{GavGF87}.}. Following this method we see that the effective
perturbation theory with respect to the radiative interaction for the matrix
elements of the scattering processes and another one for the expectation
values differ by the type of the one-particle Green function due to the
nontrivial difference between a final vacuum, $|0,out>$, and an initial
vacuum, $|0,in>$, $\;c_{v}=<0,out|0,in>$, $\left| c_{v}\right| ^{2}\neq 1$.$%
\;$ Feynman diagrams for the matrix elements of the scattering processes
have to be calculated by means of the causal propagator 
\begin{equation}
\ S^{c}(x,x^{\prime })=c_{v}^{-1}i<0,out|T\psi (x)\bar{\psi}(x^{\prime
})|0,in>,\;\;  \label{w23}
\end{equation}
where $\psi (x)$ is a massive ($m$) quantum spinor field satisfying the
Dirac equation with an external field. In the calculation of the expectation
values one has to use the one-particle Green functions 
\begin{eqnarray}
S_{in}^{c}(x,x^{\prime }) &=&i<0,in|T\psi (x)\bar{\psi}(x^{\prime })|0,in>, 
\notag \\
S_{out}^{c}(x,x^{\prime }) &=&i<0,out|T\psi (x)\bar{\psi}(x^{\prime
})|0,out>.  \label{t2}
\end{eqnarray}
Both differ from the causal propagator (\ref{w23}). Additionally, these
distinct Green functions are used to represent various matrix elements of
operators of the current and energy-momentum tensor (EMT), and effective
action beginning with zeroth order with respect to radiative interaction.
Euler and Heisenberg's one-loop effective action $Y_{out-in}$ is related to
the causal propagator, $Y_{out-in}=i\mathrm{Tr}\ln S^{c}$. Varying the $%
Y_{out-in}$, given by the Fock-Schwinger proper time representation \cite%
{S51}, one gets the following matrix elements of the operators of a current
density, $j_{\mu }$, and EMT, $T_{\mu \nu }$, in the one-loop approximation: 
\begin{equation}
<j_{\mu }>^{c}=<0,out|j_{\mu }|0,in>c_{v}^{-1}\;,\;<T_{\mu \nu
}>^{c}=<0,out|T_{\mu \nu }|0,in>c_{v}^{-1}\;,  \label{t4}
\end{equation}
where the operators $j_{\mu }$ and $T_{\mu \nu }$ are in the generalized
Furry representation, 
\begin{eqnarray}
j_{\mu } &=&\frac{q}{2}\left[ \bar{\psi}(x),\gamma _{\mu }\;\psi (x)\right]
,\;\;T_{\mu \nu }=\frac{1}{2}\left( T_{\mu \nu }^{can}+T_{\nu \mu
}^{can}\right) ,  \notag \\
T_{\mu \nu }^{can} &=&\frac{1}{4}\left\{ \left[ \bar{\psi}(x),\gamma _{\mu
}P_{\nu }\;\psi (x)\right] +\left[ P_{\nu }^{\ast }\bar{\psi}(x),\gamma
_{\mu }\;\psi (x)\right] \right\} \;,  \notag \\
P_{\mu } &=&i\partial _{\mu }-qA_{\mu }(x),\;\;q=-e.  \label{t5}
\end{eqnarray}
On the other hand, at a time instant $x^{0}$ the average values of the $%
j_{\mu }$ and $T_{\mu \nu }$ operators in the one-loop approximation are the
following 
\begin{equation}
<j_{\mu }>^{in}=<0,in|j_{\mu }|0,in>,\;<T_{\mu \nu }>^{in}=<0,in|T_{\mu \nu
}|0,in>.  \label{t6}
\end{equation}
The equalities $<j_{\mu }>^{in}=<j_{\mu }>^{c}$and $<T_{\mu \nu
}>^{in}=<T_{\mu \nu }>^{c}$ hold strictly for theory with stable vacuum.
Thus, the well known explicit expression of the $Y_{out-in}$ \cite{S51} is
useless for the calculation of any average values and we see it is desirable
to find the relevant one-loop effective description for QED with a constant
uniform electromagnetic field.

\section{Proper time representation}

To see the difference between $S_{in}^{c}$and $S^{c}$ explicitly one can
express these functions via the sets of the appropriate solutions of the
Dirac equation in an external field (see details in \cite{FGS91}). First, we
need two complete and orthonormal sets of the in/out-solutions of the Dirac
equation, $\left\{ _{\pm }\psi _{n}(x)\right\} /\left\{ ^{\pm }\psi
_{n}\left( x\right) \right\} $. They describe particles ($+$) and
antiparticles ($-$) at the initial/final time instant $%
x_{in}^{0}/x_{out}^{0} $. Second, we find decomposition coefficients $%
G\left( {}_{\zeta }|{}^{\zeta ^{\prime }}\right) $ of the out-solutions in
the in-solutions \footnote{%
We are using a convention of summation/integration over discrete/continuous
repeated indices and a compact notation where all summations/integrations
are supressed, for example $\psi _{n}G_{nm}=\left( \psi G\right) _{m}$. In
addition 
h{\hskip-.2em}\llap{\protect\rule[1.1ex]{.325em}{.1ex}}{\hskip.2em}%
=c=1 throughout this paper.}, 
\begin{equation}
{}^{\zeta }\psi (x)={}_{+}\psi (x)G\left( {}_{+}|{}^{\zeta }\right)
+\;{}_{-}\psi (x)G\left( {}_{-}|{}^{\zeta }\right) \;,  \label{emt18}
\end{equation}
Then from (\ref{w23}) we get the Feynman definition: 
\begin{eqnarray}
S^{c}\left( x,x^{\prime }\right) &=&\theta \left( x_{0}-x_{0}^{\prime
}\right) S^{-}\left( x,x^{\prime }\right) -\theta \left( x_{0}^{\prime
}-x_{0}\right) S^{+}\left( x,x^{\prime }\right) ,  \notag \\
S^{-}\left( x,x^{\prime }\right) &=&i\sum_{n,m}{}^{+}\psi _{n}\left(
x\right) G\left( \left. _{+}\right| ^{+}\right) _{nm}^{-1}\smallskip \ _{+}%
\bar{\psi}_{m}\left( x^{\prime }\right) ,  \notag \\
S^{+}\left( x,x^{\prime }\right) &=&i\sum_{n,m}\smallskip \ _{-}\psi
_{n}\left( x\right) \left[ G\left( \left. _{-}\right| ^{-}\right) ^{-1}%
\right] _{nm}^{\ast }\smallskip \ ^{-}\bar{\psi}_{m}\left( x^{\prime
}\right) ,  \label{emt19}
\end{eqnarray}
and for $S_{in}^{c}$ we have 
\begin{eqnarray}
S_{in}^{c}\left( x,x^{\prime }\right) &=&\theta \left( x_{0}-x_{0}^{\prime
}\right) S_{in}^{-}\left( x,x^{\prime }\right) -\theta \left( x_{0}^{\prime
}-x_{0}\right) S_{in}^{+}\left( x,x^{\prime }\right) ,  \notag \\
S_{in}^{\mp }\left( x,x^{\prime }\right) &=&i\sum_{n}\smallskip \ _{\pm
}\psi _{n}\left( x\right) _{\pm }\bar{\psi}_{n}\left( x^{\prime }\right) .
\label{t7}
\end{eqnarray}
Then one can express the difference as follows, 
\begin{eqnarray}
S^{a}(x,x^{\prime }) &=&S^{c}(x,x^{\prime })-\ S_{in}^{c}(x,x^{\prime }), 
\notag \\
S^{a}(x,x^{\prime }) &=&-i\sum_{nm}\smallskip \ _{-}{\psi }_{n}(x)\,\left[
G(_{+}|^{-})G(_{-}|^{-})^{-1}\right] _{nm}^{\dagger }{_{+}\bar{\psi}}%
_{m}(x^{\prime })\quad ,  \label{emt21}
\end{eqnarray}
and the similar expression can be written for $S^{p}(x,x^{\prime
})=S^{c}(x,x^{\prime })-\ S_{out}^{c}(x,x^{\prime }).$ Only if the vacuum is
stable then all the coefficients $G(_{+}|^{-})$, and then $S^{a}$, $S^{p}$
are equal to zero.

We consider the general case of a constant uniform electromagnetic field, $%
F_{\mu \nu }$, with nonzero invariants where an electric field is given by
the time dependent potential. For simplicity, we choose the reference frame
in which the electric, $\mathbf{E}$, and magnetic, $\mathbf{B}$, fields are
parallel and directed along the $x^{3}$ axis.

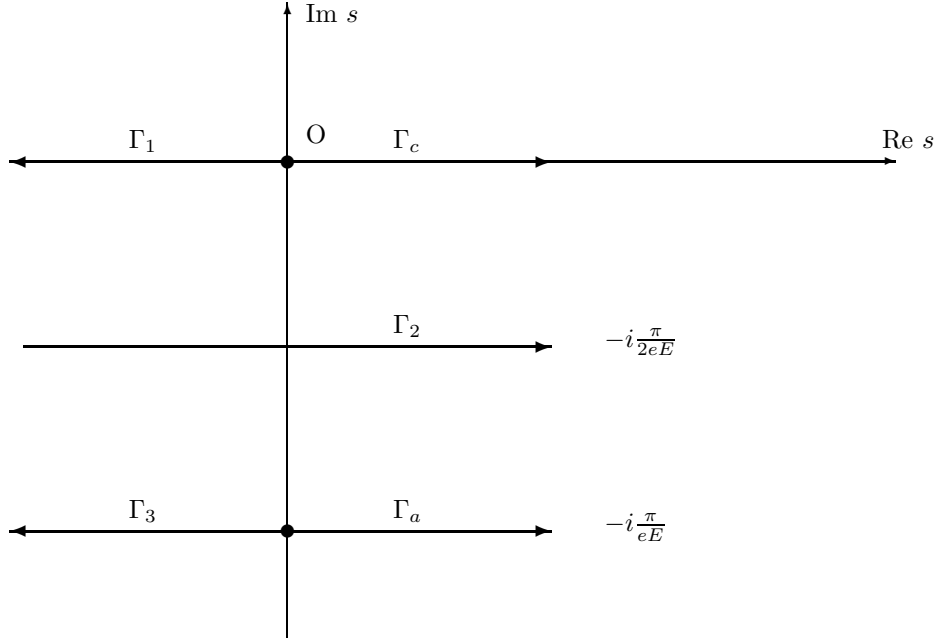
\begin{figure}[h]
\begin{picture}(350,245)
\put(0,180){\vector(1,0){330}}
\put(325,185){Re $s$}
\put(100,0){\vector(0,1){240}}
\put(107,233){Im $s$}
{\thicklines \put(0,180){\vector(1,0){200}}
\put(0,110){\vector(1,0){200}} \put(0,40){\vector(1,0){200}}
\put(0,180){\vector(-1,0){5}} \put(0,40){\vector(-1,0){5}} }
\put(140,185){$\Gamma_c$} \put(140,115){$\Gamma_2$}
\put(140,45){$\Gamma_a$} \put(40,185){$\Gamma_1$}
\put(40,45){$\Gamma_3$} \put(220,110){$-i\frac{\pi}{2eE}$}
\put(220,40){$-i\frac{\pi}{eE}$}
\put(100,180){\circle*{5}}
\put(100,40){\circle*{5}}
\put(107,187){O}
\end{picture}
\caption[f3]{{Contours of integration $\Gamma _{1},\Gamma _{2},\Gamma
_{3},\Gamma _{c},\Gamma _{a}$}.}
\label{f3}
\end{figure}

All the singular functions in a constant uniform electromagnetic field can
be represented \cite{GGG98} as the following Fock-Schwinger proper time
integrals,

\begin{eqnarray}
S^{c,a,p}(x,x^{\prime }) &=&(\gamma P+m)\Delta ^{c,a,p}(x,x^{\prime }), 
\notag \\
\Delta ^{c}(x,x^{\prime }) &=&\int_{\Gamma _{c}}f(x,x^{\prime
},s)ds=\int_{0}^{\infty }f(x,x^{\prime },s)ds,  \notag \\
\Delta ^{a/p}(x,x^{\prime }) &=&\frac{1}{2}\Delta ^{\Gamma _{2}}(x,x^{\prime
})+\Delta ^{\bar{a}/\bar{p}}(x,x^{\prime }),  \notag \\
\Delta ^{\Gamma _{2}}(x,x^{\prime }) &=&\int_{\Gamma _{2}}f(x,x^{\prime
},s)ds,  \notag \\
\Delta ^{\bar{a}/\bar{p}}(x,x^{\prime }) &=&\left[ \Theta (\pm y_{3})-\frac{1%
}{2}\right] \int_{\Gamma _{2}}f(x,x^{\prime },s)ds+\int_{\Gamma
_{a}}f(x,x^{\prime },s)ds\;  \notag \\
+ &&\Theta (\pm y_{3})\int_{\Gamma _{3}-\Gamma _{a}}f(x,x^{\prime
},s)ds\;,\;\;y_{3}=x_{3}-x_{3}^{\prime }  \label{emt31}
\end{eqnarray}
where $f(x,x^{\prime },s)$ is the known Fock-Schwinger proper time kernel %
\cite{S51} and all the contours of the integrals are shown on Fig. \ref{f3}.
The contours $\Gamma _{c}$ and $\Gamma _{1}$ are placed below the singular
points on the real axis everywhere outside of the origin. Outside of the
origin the kernel has only one singular point, $s_{1}=-i\pi /eE$, on the 
\noindent complex region between the line of the contours $\Gamma
_{c}-\Gamma _{1}$ and the line of the contours $\Gamma _{a}-\Gamma _{3}.$

By using these representations one can uniformly express all the matrix
elements of the $j_{\mu }$ and $T_{\mu \nu }$ operators, as follows 
\begin{eqnarray}
&<&j_{\mu }>^{in}=<j_{\mu }>^{c}-<j_{\mu }>^{a},\;<T_{\mu \nu
}>^{in}=<T_{\mu \nu }>^{c}-<T_{\mu \nu }>^{a},  \notag \\
&<&j_{\mu }>^{out}=<j_{\mu }>^{c}-<j_{\mu }>^{p},\;<T_{\mu \nu
}>^{out}=<T_{\mu \nu }>^{c}-<T_{\mu \nu }>^{p},  \notag \\
&<&j_{\mu }>^{c,a,p}=iq\left. \mathrm{tr}_{s}\left\{ \gamma _{\mu }\gamma
^{\nu }P_{\nu }\Delta ^{c,a,p}(x,x^{\prime })\right\} \right| _{x=x^{\prime
}}\;,  \notag \\
&<&T_{\mu \nu }>^{c,a,p}=i\left. \mathrm{tr}_{s}\left\{ B_{\mu \nu }\Delta
^{c,a,p}(x,x^{\prime })\right\} \right| _{x=x^{\prime }}\;,  \notag \\
&&B_{\mu \nu }=1/4\left\{ \gamma _{\mu }\left( P_{\nu }+P\mathcal{^{\prime }}%
_{\nu }^{\ast }\right) +\gamma _{\nu }\left( P_{\mu }+P\mathcal{^{\prime }}%
_{\mu }^{\ast }\right) \right\} \gamma ^{\kappa }P_{\kappa }\;,\;\,  \notag
\\
&&P_{\mu }^{\prime \ast }=-i\frac{\partial }{\partial x^{\prime }{}^{\mu }}%
-q\,A_{\mu }(x^{\prime }),  \label{emt24}
\end{eqnarray}
where $\mathrm{tr}_{s}\left\{ \ldots \right\} $ is the trace of an product
of the Dirac gamma matrices.

The expression for the term $<j_{\mu }>^{c}$ in (\ref{emt24}) is finite
after the proper time regularization lifting and equal to zero. The
components $<j_{\mu }>^{a/p}$ for $\mu \neq 3$ are equal to zero, as well.
All the off-diagonal matrix elements of $<T_{\mu \nu }>^{c,a,p}$ are equal
to zero. It is precisely the term $<T_{\mu \nu }>^{c}$ that can be derived
from the Heisenberg-Euler effective Lagrangian, $\mathcal{L}$. Performing
the standard renormalizations, leaving $eF_{\mu \nu }$ invariant, one gets
the finite expression of the $<T_{\mu \nu }>^{c}$ as follows 
\begin{eqnarray*}
&<&T_{00}>_{eff}^{c}=-<T_{33}>_{eff}^{c}=E\frac{\partial \mathcal{L}}{%
\partial E}-\mathcal{L},\;\;<T_{11}>_{eff}^{c}=<T_{22}>_{eff}^{c}=\mathcal{L-%
}B\frac{\partial \mathcal{L}}{\partial B}, \\
&&\mathcal{L}=\int_{0}^{\infty }\frac{ds}{8\pi ^{2}s}e^{-im^{2}s}\left[
e^{2}EB\coth \left( eEs\right) \cot \left( eBs\right) -\frac{1}{s^{2}}-\frac{%
e^{2}}{3}\left( E^{2}-B^{2}\right) \right] .
\end{eqnarray*}

Bearing in mind that $<T_{\mu \nu }>^{\Gamma _{2}}=2i\mbox{Im}<T_{\mu \nu
}>^{c},$ we get for the average values of the operators $j_{\mu }$ and $%
T_{\mu \nu }$ the following explicitly real expressions, 
\begin{equation}
<j_{\mu }>^{in}=-<j_{\mu }>^{\bar{a}},\;<T_{\mu \nu }>_{eff}^{in}=\mbox{Re}%
<T_{\mu \nu }>_{eff}^{c}-<T_{\mu \nu }>^{\bar{a}}.  \label{t9}
\end{equation}
The terms $<j_{\mu }>^{\bar{a}/p}$ and $<T_{\mu \nu }>^{\bar{a}/p}$ are
proportional to the factor $\exp \left\{ -\pi m^{2}/eE\right\} $. Thus, they
are related to global features of the theory and indicate the vacuum
instability. These matrix elements are free from the standard ultraviolet
divergences. However, with such terms we run into special kind of
divergences in the constant electric field due to the contributions from
derivatives of $\Theta (\pm y_{3})$ functions and singular point $s_{1}$.
The nature of such special divergences is connected with infinite time $T$
of acting of the constant electric field. They have to be regularized with
respect to time $T$ of acting of a constant electric field.

\section{Finite work regularization}

The state of the quantum system in question is far-from-equilibrium due to
the influence of the time dependent potential of an electric field. Then
there exists the problem of time dependence for average values which we
discuss here. In a physically correct statement of the problem, we only
refer to a quasiconstant electric field which is effectively acting during a
finite time $T$, $E(x^{0})=E$ for $t_{1}\leq x^{0}\leq t_{2}$, $%
t_{2}=-t_{1}=T/2$, and then does finite work in a finite volume. Out of the
time interval $T$ an electric field is absent. Further we call it a $T$%
-constant field. In this case the initial vacuum is the vacuum of free
particles. General aspects of the special regularization with respect to
time $T$ \ by using the $T$-constant field was discussed in \cite{GavG96a}.
Now, we need to apply those results for calculating of the leading terms in $%
<j_{3}>^{a/p}$ and $<T_{\mu \nu }>^{a/p}$at $T\rightarrow \infty $.

The mean number of particles created by the external field from the initial
vacuum is 
\begin{equation}
R_{n}^{cr}=<0,in|a_{n}^{\dagger }(out)a_{n}(out)|0,in>=\left| G\left(
{}_{-}|{}^{+}\right) \right| ^{2},  \label{e11}
\end{equation}%
were the standard volume regularization was used, so that $\delta (\mathbf{p}%
-\mathbf{p}^{\prime })\rightarrow \delta _{\mathbf{p},\mathbf{p}^{\prime }}$%
. If the time $T$ is sufficiently large: $T>>T_{0}$, where $T_{0}=(1+\lambda
)/\sqrt{eE}$ is called the stabilization time, and $eET/2\gg |p_{3}|$, then 
\begin{eqnarray}
R_{n}^{cr} &=&e^{-\pi \lambda }\left[ 1+O\left( \left[ \frac{1+\lambda }{K}%
\right] ^{3}\right) \right] ,\;\;-\sqrt{eE}\frac{T}{2}\leq \xi \leq -K,
\label{e12} \\
\lambda &=&\frac{m^{2}+\left\langle P_{\bot }^{2}\right\rangle }{eE}%
,\;P_{\bot }=(P^{1},P^{2},0),\;\xi =(\left| p_{3}\right| -eET/2)/\sqrt{eE}, 
\notag
\end{eqnarray}%
where $K$ is a sufficiently large arbitrary constant, $K>>1+\lambda $, $%
\left\langle P_{\bot }^{2}\right\rangle $ is the conserved average value of
the $P_{\perp }^{2}$, and $p_{3}$ is a longitudinal momentum. The $%
R_{n}^{cr} $ distribution for large longitudinal momenta, $|p_{3}|>>eET/2$,
decreases, $R_{n}^{cr}=O\left( \left[ \lambda /\xi ^{2}\right] ^{3}\right) $%
. The latter expression allows one to consider the limit $T\rightarrow
\infty $ at any given quantum number. In this limit the distribution
function takes the simple form $R_{n}^{cr}=e^{-\pi \lambda }\;$which
coincides with the expressions obtained in the constant electric field \cite%
{Nik70}.

The distribution $R_{m}^{cr}$ plays role of the cut-off factor for the
integral (\ref{emt21}) and similar representation of the $S^{p}$, then the
contributions of $S^{a/p}$ are convergent. If the time interval $%
x^{0}-t_{1}=x^{0}+T/2$ is sufficiently large, $\sqrt{eE}\left(
x^{0}+T/2\right) \gg 1+m^{2}/eE$, we can extract the leading contributions
at large $T$ (marked a subscribt ''as'') in the representations (\ref{emt24}%
), (\ref{t9}) and then, integrating over quantum numbers and calculating
derivatives, obtain that 
\begin{eqnarray}
&<&j_{\mu }>_{as}^{a/p}=-\delta _{\mu }^{3}2e\left( 1/2\pm x^{0}/T\right)
n^{cr},  \notag \\
&<&T_{00}>_{as}^{a/p}=<T_{33}>_{as}^{a/p}=-eET\left( 1/2\pm x^{0}/T\right)
^{2}n^{cr},  \notag \\
&<&T_{11}>_{as}^{a/p}=<T_{22}>_{as}^{a/p}  \notag \\
&=&\widetilde{n}\left\{ 
\begin{array}{c}
\mp \ln \left[ \sqrt{eE}\left( T/2\pm x^{0}\right) \right] +O\left( \ln
K\right) \;\mathrm{if\;}\sqrt{eE}\left( T/2\pm x^{0}\right) >K \\ 
O\left( \ln K\right) \;\mathrm{if\;}\sqrt{eE}\left( T/2\pm x^{0}\right) \leq
K%
\end{array}
\right. ,  \label{emt44}
\end{eqnarray}
where $K$ is an arbitrary constant, $K\gg 1+m^{2}/eE$, 
\begin{eqnarray}
n^{cr} &=&\frac{e^{2}EBT}{4\pi ^{2}}\coth \frac{\pi B}{E}\left[ \exp \left\{
-\pi \frac{m^{2}}{eE}\right\} +O\left( \frac{K}{\sqrt{eE}T}\right) \right]
\smallskip ,  \notag \\
\widetilde{n} &=&\frac{e^{2}B^{2}}{4\pi ^{2}\sinh ^{2}\left( \pi B/E\right) }%
\exp \left\{ -\pi \frac{m^{2}}{eE}\right\} .\;\;  \label{emt45}
\end{eqnarray}
Note that here $n^{cr}$ is a characteristic number density of excitable
states in the external field and, as we see subsequently, it is the same as
the number density of the created pairs for time $T$ of the duration of the
electric field.

The current density and the EMT of the final particles created from vacuum
by the $T$-constant field for the large time interval $x^{0}-t_{1}=x^{0}+T/2%
\gg K/\sqrt{eE}$ can be presented as 
\begin{equation}
j_{\mu }^{cr}=<j_{\mu }>^{in}-<j_{\mu }>^{out}\;,\;\;T_{\mu \nu
}^{cr}=<T_{\mu \nu }>^{in}-<T_{\mu \nu }>^{out}\;,\;  \label{emt46}
\end{equation}
where the terms $<j_{\mu }>^{out}$ and $<T_{\mu \nu }>^{out}$ are used to
take into account the normal ordering of the current density and the EMT
operators with respect to creation and annihilation operators of the final
particles. Then one gets from (\ref{emt24}) and (\ref{emt44}) that 
\begin{eqnarray}
j_{\mu }^{cr} &=&<j_{\mu }>_{as}^{p}-<j_{\mu }>_{as}^{a}=\delta _{\mu
}^{3}2e\left( 2x^{0}/T\right) n^{cr},  \notag \\
T_{\mu \nu }^{cr} &=&<T_{\mu \nu }>_{as}^{p}-<T_{\mu \nu }>_{as}^{a},
\label{emt47}
\end{eqnarray}
and 
\begin{eqnarray}
&<&T_{00}>^{cr}=<T_{33}>^{cr}=2eEx^{0}n^{cr},  \label{emt48} \\
&<&T_{11}>^{cr}=<T_{22}>^{cr}  \notag \\
&=&\widetilde{n}\left\{ 
\begin{array}{c}
\ln \left[ eE\left( \left( T/2\right) ^{2}-\left( x^{0}\right) ^{2}\right) %
\right] +O\left( \ln K\right) \;\mathrm{if\;}\sqrt{eE}\left(
T/2-x^{0}\right) >K \\ 
\ln \left[ \sqrt{eE}\left( T/2+x^{0}\right) \right] +O\left( \ln K\right) \;%
\mathrm{if\;}\sqrt{eE}\left( T/2-x^{0}\right) \leq K%
\end{array}
\right. .  \notag
\end{eqnarray}
At $x_{0}=t_{2}=T/2$ one gets from (\ref{emt47}), (\ref{emt48}) the
expressions for the total current densities and the EMT of the particles
created.

\section{Conclusion}

We finally obtain the average values of the current density and the EMT as
following 
\begin{equation}
<j_{\mu }>^{in}=-<j_{\mu }>_{as}^{a},\;<T_{\mu \nu }>_{eff}^{in}=\mbox{Re}%
<T_{\mu \nu }>_{eff}^{c}-<T_{\mu \nu }>_{as}^{a}.  \label{emt52}
\end{equation}
As we have seen, the $T$ dependent contributions to $<j_{\mu }>^{\bar{a}}$
and $<T_{\mu \nu }>^{\bar{a}}$ appear due to the vacuum instability and then
come with the factor $\exp \left\{ -\pi m^{2}/eE\right\} $. This factor is
exponentially small for a weak electric field, $m^{2}/eE\gg 1$, and the
effect can actually be observed as soon as the external field strength
approaches the characteristic value $E_{c}=m^{2}/e$. On the other hand, the
term $\mbox{Re}<T_{\mu \nu }>_{eff}^{c}$ is the $T$ independent and its
contribution is not small whether the electric field is weak or strong. When
the T-constant electric field is switched off at $x^{0}>T/2$, the local
vacuum contribution of the $E$ in the $\mbox{Re}<T_{\mu \nu }>_{eff}^{c}$ is
absent but the global contribution given by the $\left. <T_{\mu \nu
}>_{as}^{a}\right| _{x^{0}=T/2}$ is presence. Thus, in general case both
kinds of contributions are important.

Using the expression (\ref{emt52}) we find a condition for validity of a
strong constant electric field concept. With a very strong $E$ field, $%
m^{2}/eE\ll 1$ ($B=0$), and large $T$ one gets the well known asymptotic
expression of the $\mbox{Re}<T_{00}>_{eff}^{c}$ vacuum energy density, 
\begin{equation*}
\mbox{Re}<T_{00}>_{eff}^{c}=-\frac{e^{2}}{24\pi ^{2}}E^{2}\ln \frac{eE}{m^{2}%
}.
\end{equation*}%
It is $T$ independent contribution. The energy density of a classic electric
field is $E^{2}/8\pi $. Then it seems that an electric field concept is
physically meaningful when $\frac{e^{2}}{3\pi }\ln \frac{eE}{m^{2}}\ll 1$.
But when $T$ \ is large one has to give attention to the $T$ dependent term $%
<T_{00}>_{as}^{a}$. At $x^{0}=T/2$ we have 
\begin{equation*}
-<T_{00}>_{as}^{a}=\frac{e^{2}E^{2}}{4\pi ^{3}}eET^{2}.
\end{equation*}%
Of course, one can neglect a back-reaction on an electric field only if the
last term is far less than $E^{2}/8\pi $. Thus, the true condition for
validity of a strong constant electric field concept is the following 
\begin{equation*}
1\ll eET^{2}\ll \frac{\pi ^{2}}{2e^{2}}.
\end{equation*}

All the results for a pair creation are valid within the accuracy of the
analysis at low density and temperature, $\Theta \ll eET$.

\end{document}